\documentclass[english,twocolumn,english,aps, prl, reprint, numbers, square, comma, sort&compress, showpacs]{revtex4-1}
\usepackage[T1]{fontenc}
\usepackage[latin9]{inputenc}
\usepackage{amsmath}
\usepackage{graphicx}
\usepackage[dvipdfm,colorlinks=true, linkcolor=blue, citecolor=blue, urlcolor=blue]{hyperref}



\usepackage{babel}

\begin{document}

\title{Theory of light amplification in active fishnet metamaterials}

\author{Joachim M. Hamm, Sebastian Wuestner, Kosmas L. Tsakmakidis, and Ortwin
Hess}

\email{o.hess@imperial.ac.uk}

\affiliation{Department of Physics, South Kensington Campus, Imperial College,
London SW7 2AZ, UK}
\begin{abstract}
We establish a theory that traces light amplification in an active
double-fishnet metamaterial back to its microscopic origins. Based
on \emph{ab initio} calculations of the light/plasmon fields we extract
energy rates and conversion efficiencies associated with gain/loss
channels directly from Poynting's theorem. We find that for the negative
refactive index mode both radiative loss and gain outweigh resistive
loss by more than a factor of two, opening a broad window of steady-state
amplification (free of instabilities) accessible even when a gain
reduction close to the metal is taken into account.
\end{abstract}

\maketitle
Recent advances in metamaterials research and active plasmonics bring
with them the prospect for next generation, loss-free optical metamaterial
designs \citep{Cai_2009}. The double-fishnet metamaterial, first demonstrated in the
near infrared \citep{Zhang_2005,*Zhang_2005_2}, is a simple nanoplasmonic
structure with a broadband effective negative refractive index, the
origins of which are now well established \citep{Mary_2008}. To date
it remains the prime example of a metamaterial able to operate at
visible wavelengths \citep{Valentine_2008} and serves as a template
for novel optical metamaterial designs \citep{Soukoulis_2011}. In
plasmonics an understanding of how losses experienced by surface-plasmon
polaritons (SPPs) and localized plasmons can be overcome by gain media
placed adjacent to metals \citep{Ramakrishna_2003,Plum_2009,Meinzer_2010} has had
far reaching impact. This is best illustrated by the realization of
SPP amplifiers \citep{DeLeon_2010} and nano-plasmonic lasers \citep{Oulton_2009}
but also by recent developments in optical metamaterials research.
In 2010, first evidence has been given that loss compensation in active
plasmonic metamaterials is possible: Experimental \citep{Xiao_2010}
and theoretical \citep{Wuestner_2010,Fang_2010_2} absorption/transmission/reflection
(ATR) studies on optically pumped gain-enhanced double-fishnet metamaterials
showed that gain can overcome the internal dissipative losses while
retaining and in fact enhancing the effective negative refractive
index response. 
%Employing a mesoscopic theory that describes the double-fishnet metamaterial as a meshed 
%network of plasmonic channels Yang et.~al.~find that loss compensation
%is closely connected to compensating the damping of the gap-SPP \cite{Yang_2011}.
%%the point of loss-compensation is encountered when the gap-SPP becomes undamped \citep{Yang_2011}. 
%This phenomenological treatment of gain as an effective parameter
%does however not capture the complex interaction of the
%strong local fields of the gap-SPP with the spatially inhomogeneous inversion
%in an optically pumped plasmonic structure \citep{Sivan_2009,Wuestner_2010}.
%Ultimately, a comprehensive and self-consistent 
%theory is required to elucidate the internal energy flow, inaccessible by experiment,
%and to derive loss management strategies for the design of future
%practical metamaterials.
Further theoretical work has shown that loss compensation is achieved when the gap-SPP
becomes undamped \cite{Yang_2011} causing the resonance linewidth to be limited by
scattering losses only \cite{Wuestner_2011,Yang_2011}. The physical origin of light
amplification in active fishnet metamaterials lies in the complex 
interaction of the strong local fields of the gap-SPP mode with the spatially inhomogeneous 
inversion generated by a pump process \citep{Sivan_2009,Wuestner_2010}.
Ultimately, to develop loss-management strategies for future practical metamaterials, 
a theory is required that traces the internal energy flow, inaccessible by experiment,
on the nanometer scale.

\begin{figure}
 \includegraphics{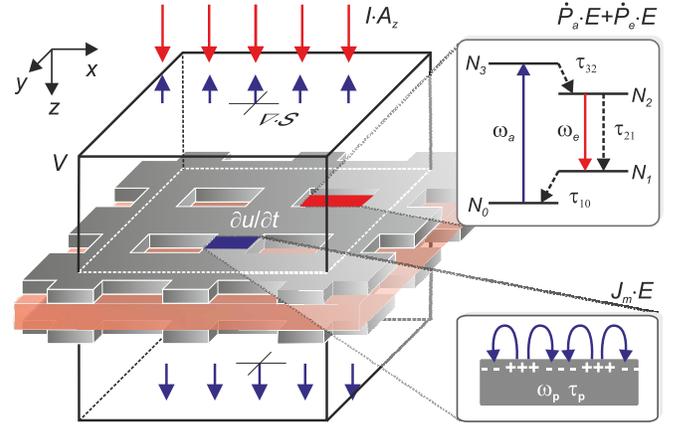}
 \caption{\label{fig:1}(color online) Schematic of an active plasmonic metamaterial
illuminated by an incident plane wave of intensity $I$ that creates
a power flux into the volume $V$ through the surface $A_{\mathrm{z}}$.
According to Poynting's theorem {[}see Eq.~(\ref{eq-1}){]} the electromagnetic
energy $U$ inside the volume changes due to energy transfer into
and out of various loss (blue) and gain channels (red). Note that
for normal incidence onto an infinite film, the flux between neighbouring
volumes cancels.}
\end{figure}

In this Letter, we present a theoretical framework that allows us
to describe the microscopic dynamical processes, which, collectively,
make light amplification in active metamaterials possible. We conduct,
under realistic assumptions, extensive numerical pump and probe studies
and extract together with the macroscopic spectral response
the effective rates of energy exchange between physical subsystems
of an active fishnet metamaterial. The results demonstrate the intricate mechanisms that
make steady-state amplification in nano-plasmonic structures possible:
the efficiency of the pump scheme affecting the spatial deposition of inversion,
the electric field enhancement that increases the effective gain, 
and the dominance of radiative coupling over internal losses.

Our studies are based on the Maxwell-Bloch (MB) approach \citep{Wuestner_2011,Bohringer_2008,*Bohringer_2008_2}
combined with a suitable formulation of Poynting's theorem. We solve
Maxwell\textquoteright{}s field equations in full-vectorial three-dimensional
form and in interaction with the dynamic polarization responses ascribed
to the metal and four-level gain medium microscopically, at every point in space,
while simultaneously tracing the flow of electromagnetic energy into
and out of microscopic channels (see Fig.~\ref{fig:1}). This generic
method allows us to dynamically record and analyze the gain and loss
rates associated with the physical subsystems of any structure and
thereby to identify transient and steady-state regimes of operation
as well as to precisely determine the conditions for amplification
and lasing.

We start by recalling Poynting's theorem, according to which the rate
of change of the electromagnetic energy inside a volume $V$ that
contains metal (loss) and four-level gain media is determined by 
\begin{equation}
\langle\partial u/\partial t\rangle=-\langle\nabla\cdot\mathbf{S}\rangle-\langle\dot{\mathbf{P}}_{\mathrm{f}}\cdot\mathbf{E}\rangle-\langle\dot{\mathbf{P}}_{\mathrm{a}}\cdot\mathbf{E}\rangle-\langle\dot{\mathbf{P}}_{\mathrm{e}}\cdot\mathbf{E}\rangle\label{eq-1}
\end{equation}
where $u(\mathbf{r},t)=\frac{1}{2}[\varepsilon_{0}\varepsilon(\mathbf{r})E^{2}(\mathbf{r},t)+\mu_{0}H^{2}(\mathbf{r},t)]$
is the energy density, $\mathbf{S}(\mathbf{r},t)$ the Poynting vector,
$\mathbf{E}(\mathbf{r},t)$ the electric field, and $\mathbf{P}_{i}(\mathbf{r},t)$
the polarizations of the free electron plasma in the metal ($i=\mathrm{f}$)
and the gain media at the transitions of absorption and emission ($i=\mathrm{a,e}$).
The operator $\langle\bullet\rangle$ performs an integration over
$V$ and a time-averaging over $\Delta t\gtrsim2\pi/\omega$ eliminating
fast phase oscillations. Equation \eqref{eq-1} expresses the change
of electromagnetic energy in $V$ as influx through the surface minus
the work that the field exercises on the electrons in the metal and
on the transitions of the gain media.

It will prove useful to write Eq.~\eqref{eq-1} as a rate equation:
\begin{equation}
\partial U/\partial t=-\Gamma_{\mathrm{t}}U=-\Lambda U-\Gamma_{\mathrm{f}}U-\Gamma_{\mathrm{a}}U-\Gamma_{\mathrm{e}}U\label{eq-3}
\end{equation}
where $U=\langle u+\sum_{i}w_{i}\rangle$ is the total field energy
in the volume, $\Gamma_{\mathrm{t}}$ the energy decay rate, $\Lambda=\left\langle \nabla\cdot\mathbf{S}\right\rangle /U$
the flux rate that measures the outflow of energy, and $\Gamma_{\mathrm{i}}=[\bigl\langle\dot{\mathbf{P}}_{\mathrm{i}}\cdot\mathbf{E}-\partial w_{i}/\partial t\bigr\rangle]/U$
the power loss/gain rates. $w_{i}$ are contributions of the dynamic
responses to the overall electromagnetic energy in the medium that
can be calculated as in \citep{Ruppin_2002}. Note that the so-defined
effective rates are time-dependent quantities and become constant
when $U$ is either constant or decreases exponentially during single mode decay.
We also emphasize that this formulation of volume-averaged
energy transfer rates is exact (since it directly reflects energy
conservation) and applicable to arbitrary linear and nonlinear systems.
\begin{figure}
\includegraphics{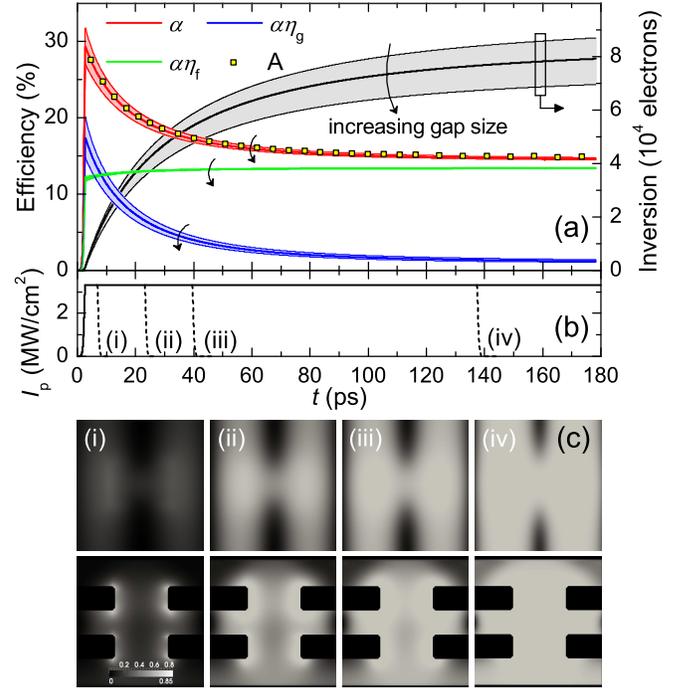}
\caption{(color online) Conversion efficiencies over pump duration $\tau_{\mathrm{p}}$
for gap sizes of $0-10\,\mathrm{nm}$ (thick lines: $5\,\mathrm{nm}$). (a) Left axis: Total absorption
efficiency $\alpha$ (red lines), external pump efficiency $\alpha\eta_{\mathrm{g}}$
(blue lines), and dissipative efficiency $\alpha\eta_{\mathrm{f}}$
(green lines); absorption coefficient $A$ (yellow squares) obtained
from ATR calculations; Right axis: Total inversion $\langle N_2 - N_1 \rangle$(black lines) of
the gain material. (b) Pump intensity $I_{\mathrm{p}}$ over $\tau_{\mathrm{p}}$
indicating the switch-off times (dashed lines). (c) Spatial inversion
profiles, $\left[N_2(\mathbf{r})-N_1(\mathbf{r})\right]/N$, for four different pump times (i)-(iv): (top row) x-y cross
section in the middle between the fishnet metal films; (bottom row)
x-z cross-section through the centers of the holes.\label{fig:2}}
\end{figure}

The object of our studies is a $260\,\mathrm{nm}$ thick fishnet
metamaterial consisting of two $40\,\mathrm{nm}$-thick silver films,
separated by $60\,\mathrm{nm}$ and embedded in an active dielectric
host of refractive index $n=1.62$ (see Fig.~\ref{fig:1}). The silver
films are perforated with rectangular holes of dimensions $a_{x}=120\,\mathrm{nm}$
and $a_{y}=80\,\mathrm{nm}$ with square periodicity $p=280\,\mathrm{nm}$.
For the gain medium (dye) we assume a density of $N=3.89\times10^{18}\,\mathrm{cm}^{-3}$,
wavelength of maximum absorption (emission) $\lambda=680\,\mathrm{nm}$
($710\,\mathrm{nm}$), and associated cross-section of $\sigma=7.45\times10^{-16}\,\mathrm{cm^{2}}$
($5.78\times10^{-16}\,\mathrm{cm^{2}}$). To model the response of
silver we use a three-pole Drude-Lorentz model \citep{McMahon_2009}. 

\begin{figure}
\includegraphics{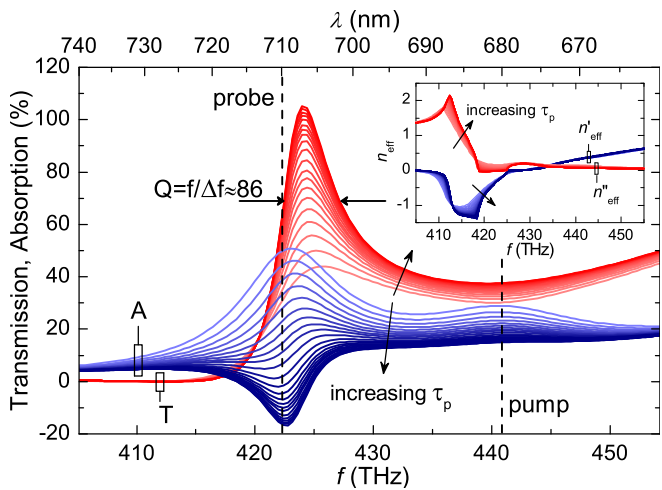}
\caption{(color online) Absorption (blue lines) and transmission spectra (red
lines) for pump times $\tau_{\mathrm{p}}=2.7-180\,\mathrm{ps}$
(lighter to darker colors, black arrows) for $5\,\mathrm{nm}$ gap
size. The $Q$-factor calculated from the width of the transmission
resonance is given for $\tau_{\mathrm{p}}=180\,\mathrm{ps}$ . Inset:
real (blue lines) and imaginary parts (red lines) of the effective
refractive index retrieved from the complex ATR spectra.\label{fig:4}}
\end{figure}

In our analysis, we first pump the active structure with a pulse of
$\lambda_{\mathrm{p}}=680\,\mathrm{nm}$ and then probe it with a
signal pulse of center-wavelength $\lambda_{\mathrm{s}}=710\,\mathrm{nm}$.
Both the pump and the probe fields are $x$-polarized, and the probe
falls within the negative-index resonance of the structure (see Fig.~\ref{fig:4} 
later on). Fabrication tolerances, passivation, and quenching
may lead to a reduction of the local gain close to metal surfaces,
which we qualitatively accommodate for by considering a gap between
the metal and the gain material with values of $g=0 - 10\,\mathrm{nm}$. 
The volume $V$ used for the rate retrieval
{[}Eq.~\eqref{eq-3}{]} is chosen to tightly enclose the structure. 

The amplification performance of the active double-fishnet critically
depends on where and how efficiently energy is deposited in the gain
medium and the metal during the pump process. While macroscopically
one can only measure the total absorption, the presented methodology
allows us to trace the conversion of energy into inversion and
thermal energy microscopically within the gain medium and the metal
films, as shown in Fig.~\ref{fig:2}. We pump the metamaterial with
a pulse of constant peak intensity $I_{\mathrm{p}}=3.3\,\mathrm{MW}/\mathrm{cm}^{2}$
but varying duration $\tau_{\mathrm{p}}=0-180\,\mathrm{ps}$ (in steps
of $1.38\,\mathrm{ps}$) and record the gradual change in the efficiency
with which energy is absorbed by the gain medium and the metal. To
this end, we define the following photon conversion efficiencies:
the total absoption efficiency $\alpha=-\langle\nabla\cdot\mathbf{S}\rangle/(I_{\mathrm{p}}A)$
is the net percentage of photons absorbed in the volume (influx);
the internal pump efficiency $\eta_{\mathrm{g}}=\Gamma_{\mathrm{g}}/\Lambda$
(with $\Gamma_{\mathrm{g}}=-\Gamma_{\mathrm{e}}-\Gamma_{\mathrm{a}}$)
is the percentage of absorbed photons transfering their energy to
the gain\emph{ }medium; and the internal dissipative {}`efficiency'
$\eta_{\mathrm{f}}=-\Gamma_{\mathrm{f}}/\Lambda$ is the percentage
of absorbed photons dissipated in the metal. 

Figure \ref{fig:2}(a) shows that just after the ramp-up of the pump
pulse ($2.7\,\mathrm{ps}$) approximately a third of all incident
photons are absorbed inside the cavity (red lines), with the efficiency
of the energy transfer to the dye (blue lines) dominating over dissipative
losses (green lines) by a factor of $\eta_{\mathrm{g}}/\eta_{\mathrm{f}}\approx1.45$.
The energy absorbed by the dye is converted into inversion, which
initially concentrates around the metal interfaces where the pump
field enhancement is strong {[}see Fig.~\ref{fig:2}(c) (i){]}. As the duration of the pump increases,
the efficiency with which inversion is created in the dye $(\alpha\eta_{\mathrm{g}})$
decreases rapidly falling below the contribution from dissipative
losses [Fig.~\ref{fig:2}(a)]. This drop in pump efficiency is caused by the saturation of
the gain medium, with areas of high inversion gradually spreading
out from close to the metal, as can be seen in Figs. \ref{fig:2}(c)
(i)$-$(iv). Eventually, a global saturation of the total inversion [Fig.~\ref{fig:2}(a), black lines] 
sets in where half the maximum value is already reached within the first $30\,\mathrm{ps}$. 

We further note that the efficiency with which energy is absorbed
by the gain medium (blue lines) does not reach zero for $\tau_{\mathrm{p}}\rightarrow\infty$
since the broad emission line of the dye overlaps with the pump wavelength.
As the medium approaches saturation the internal quantum yield decreases
and the electric field energy is increasingly absorbed by the dye
in absorption-emission cycles without creating inversion. 
\begin{figure}
\includegraphics{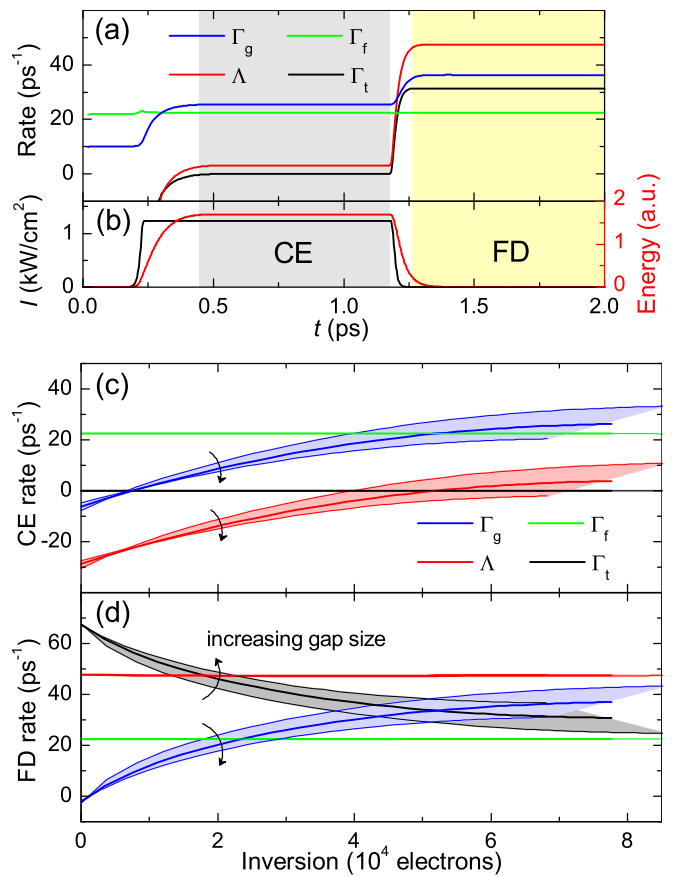}
\caption{(color online) (a) Rate dynamics during probe process for $5\,\mathrm{nm}$
gap in amplifying regime: Gain rate $\Gamma_{\mathrm{g}}$ (blue line),
dissipative loss rate $\Gamma_{\mathrm{f}}$ (green line), outflux
rate $\Lambda$ (red line), and energy decay rate $\Gamma_{\mathrm{t}}$
(black line). (b) Probe pulse intensity $I_{\mathrm{s}}$ (black line)
and energy $U$ inside the volume (red line) over time, with highlighted
regimes of continuous excitation (CE) and free decay (FD). (c,d) Extracted
rates over total inversion in the gain medium for gap sizes $0-10\,\mathrm{nm}$
under CE (c) and in FD (d) {[}colors as in (a){]}.\label{fig:3}}
\end{figure}

Having pumped the active metamaterial into inversion, we now probe
it with a signal pulse of duration $\tau_{\mathrm{s}}=1\,\mathrm{ps}$
and peak intensity $I_{\mathrm{s}}=1.23\:\mathrm{kW}/\mathrm{cm}^{2}$
(Fig.~\ref{fig:3}(b), black line) at the emission wavelength of the
dye. Note that the energy inside the resonator (Fig.~\ref{fig:3}(b),
red line) does not instantaneously follow the excitation $I_{\mathrm{s}}$,
but exhibits a rising flank during switch-on and an exponential decay
after the switch-off of the probe pulse. After the transient build-up 
of energy in the resonant mode (Fig.~\ref{fig:3}(a), negative values for the black line)
 two distinct phases of constant rates follow: steady-state continuous excitation (CE) characterized
by $\Gamma_{\mathrm{t}}=0$; and free decay (FD) of the excited mode with a $Q$-factor
$Q=\omega_{\mathrm{e}}/\Gamma_{\mathrm{t}}$. Figure \ref{fig:3}(a)
exemplifies the retrieval of rates for a specific pump duration (and
inversion) and $5\,\mathrm{nm}$ gap size. By performing corresponding
probe experiments for pump times $\tau_{\mathrm{p}}=0-180\,\mathrm{ps}$
and gap sizes $g=0-10\,\mathrm{nm}$ we obtain the results reported
in Figs. \ref{fig:3}(c) and (d).

In the CE phase {[}Fig.~\ref{fig:3}(c){]}, the energy inside the driven
resonator does not change with time and, according to Eq.~\eqref{eq-3},
the flux rate $\Lambda$ (red lines) then equals the difference between
the net gain rate $\Gamma_{\mathrm{g}}$ (blue lines) and the dissipative-loss
rate $\Gamma_{\mathrm{f}}$ (green lines), i.e. $\Lambda=\Gamma_{\mathrm{g}}-\Gamma_{\mathrm{f}}$.
Figure \ref{fig:3}(c) shows that overcoming the amplification threshold $\Lambda=0$, where the gain rate equals the loss rate, 
requires higher values of total inversion for larger gap sizes.  This is because the field of the resonator mode
at the probe wavelength $\mathbf{E}(\lambda_{\mathrm{e}},\mathbf{r})$,
which weighs the local gain rate according to $\Gamma_{\mathrm{g}}\propto\int\Gamma_{\mathrm{g}}(\mathbf{r})\left|\mathbf{E}(\lambda_{\mathrm{e}},\mathbf{r})\right|^{2}\mathrm{d}V$ \citep{Sivan_2009},
is highest close to the metal. While the inversion in these regions makes larger contributions to the effective gain,
its impact on the gain-enhancement is less strong than one might expect,
making amplification possible even for larger gap sizes of up to $g\gtrsim7.5\,\mathrm{nm}$. Energy transfer close to the metal 
interfaces (electron-hole coupling \citep{DeLeon_2010}) is therefore expected to increase the threshold but does not necessarily prevent
amplification. 

When the probe is switched off, the rates rebalance and enter the
FD phase {[}see Fig.~\ref{fig:3}(d){]}. The energy inside the resonator
then decays exponentially with a rate $\Gamma_{\mathrm{t}}=\Lambda+\Gamma_{\mathrm{f}}-\Gamma_{\mathrm{g}}>0$.
Just as the amplification threshold was characterized by $\Lambda=0$
in the CE regime, we identify the lasing threshold as $\Gamma_{\mathrm{t}}=0$
in the FD regime. Here, for lasing to occur the gain would have to
overcome the sum of dissipative and radiative loss. 

The constancy of the dissipative loss rate ($\Gamma_{\mathrm{f}}\approx22\,\mathrm{ps}^{-1}$)
and the radiative damping rate in free decay ($\Lambda\approx48\,\mathrm{ps}^{-1}$)
show that neither the overlap of the field with the metal nor the
coupling to the radiative continuum are affected by gap size or inversion.
This indicates that the mode profiles in the two regimes (CE and
FD) remain constant. Interestingly, the dissipative loss rate also remains
unaltered during the $\mathrm{CE}\rightarrow\mathrm{FD}$ transition,
despite the switch-off of the incident field. In effect, a window
of amplification exists for the gain rate $\Gamma_{\mathrm{g}}$ between the point of loss-compensation and the lasing threshold,
 $\Gamma_{\mathrm{f}}<\Gamma_{\mathrm{g}}<\Gamma_{\mathrm{f}}+\Lambda^{\mathrm{FD}}$.
For the considered active plasmonic metamaterial the radiative flux
rate $\Lambda^{\mathrm{FD}}$ (Fig.~\ref{fig:3} (d), red lines) is
more than two times larger than the dissipative-loss rate $\Gamma_{\mathrm{f}}$
(green lines). In practice, the regime of amplification might be further limited by
gain saturation as $\Gamma_{\mathrm{g}}\leq\Gamma_{\mathrm{g}}^{\mathrm{sat}}$.
For sufficiently high gain densities the lasing threshold can be crossed. The energy inside
the resonator then increases exponentially ($\Gamma_{\mathrm{t}}<0$,
lasing instability) until gain depletion sets in, which is indeed observed.

The microscopically calculated cavity $Q-$factor and the amplification
threshold can be related to the ATR spectra depicted in Fig.~\ref{fig:4}.
We see that for longer pump durations (increased inversions) all spectra
converge due to gain saturation, including those of the extracted effective refractive index \citep{Smith_2002}
shown in the inset. No further increase (decrease)
in the transmission (absorption) of the probe pulse through the structure
is possible. The width of the transmission resonance narrows to $\Delta f\approx4.9\,\mathrm{THz}$
at saturation, which is in good agreement with the extracted decay
rate of $\Gamma_{\mathrm{t}}\approx31\,\mathrm{ps}^{-1}\approx2\pi\Delta f$
{[}last point of full black line in Fig.~\ref{fig:3}(d){]}. Further, the absorption coefficient $A$ at the probe 
wavelength becomes zero at precisely the same inversion value at which $\Lambda=0$ in the CE regime {[}full
red line in Fig.~\ref{fig:3}(c){]}. At the pump wavelength, $A$ remains positive, decreases with higher
values of inversion and, presented as yellow squares in Fig.~\ref{fig:2},
agrees quantitatively with the microscopically calculated absorption
efficiency $\alpha$. 

In conclusion, we have introduced a generic Maxwell Bloch-based methodology that
allowed us to investigate the dynamic origins of light amplification,
i.e. the transfer of energy between microscopic gain and loss channels,
in a fishnet metamaterial. The extracted rates associated with the
subsystems of dye, metal, and radiative continuum reveal that radiative
damping dominates over the dissipative loss. Our results show that steady-state
net amplification free of lasing instabilities is possible over a
broad parameter regime and may be achieved even when
gain passivation close to the metal is considered.

We gratefully acknowledge support by the Leverhulme Trust, the Royal Academy of Engineering and the EPSRC (UK).

%\bibliographystyle{apsrev4-1}
%\bibliography{paper4}

%merlin.mbs apsrev4-1.bst 2010-07-25 4.21a (PWD, AO, DPC) hacked
%Control: key (0)
%Control: author (72) initials jnrlst
%Control: editor formatted (1) identically to author
%Control: production of article title (-1) disabled
%Control: page (0) single
%Control: year (1) truncated
%Control: production of eprint (0) enabled
%

\end{document}